\documentclass[prd,aps]{revtex4}
\begin{document}
\title{Remarks on a Chern-Simons-like coupling}

\author{Patricio Gaete}\email{patricio.gaete@usm.cl}
\affiliation{Departmento de F\'{i}sica and Centro Cient\'{i}fico-
Tecnol\'ogico de Valpara\'{i}so, Universidad T\'{e}cnica Federico
Santa Mar\'{i}a, Valpara\'{i}so, Chile}

\date{\today}

\begin{abstract}
We consider the static quantum potential for a gauge theory
which includes a light massive vector field interacting with the
familiar $U(1)_{QED}$ photon via a Chern-Simons- like coupling,
by using the gauge-invariant, but path-dependent, variables formalism.
An exactly screening phase is then obtained, which displays
a marked departure of a qualitative nature from massive axionic
electrodynamics. The above static potential profile is
similar to that encountered in axionic electrodynamics consisting of
a massless axion-like field, as well as to that encountered in the
coupling between the familiar $U(1)_{QED}$ photon and a second
massive gauge field living in the so-called $U(1)_h$ hidden-sector,
inside a superconducting box.
\end{abstract}

\maketitle

\section{Introduction}

The existence of axion-like particles and light extra hidden $U(1)$
gauge bosons have been proposed in many investigations of extensions
of the Standard Model (SM), in order to explain cosmological and
astrophysical results $^{1-13}$. Let us recall here that the axion-like
scenario can be qualitatively  understood by the existence of light
pseudoscalar bosons $\phi$ ("axions"), with a coupling to two photons.
In this case the interaction term in the effective Lagrangian has the
form $\mathcal{L}_I  =  - \frac{1}{4}F_{\mu \nu } \tilde F^{\mu \nu } \phi$,
where $\tilde F^{\mu \nu }  = \frac{1}{2}\varepsilon _{\mu \nu \lambda
\rho} F^{\lambda \rho }$. However, the crucial feature of axionic
electrodynamics is the mass generation due to the breaking of rotational
invariance induced by a classical background configuration of the gauge
field strength \cite{14}, which leads to confining potentials in the
presence of nontrivial constant expectation values for the gauge field
strength \cite{15}.

In this perspective, in recent times a different extension of the SM has
been considered \cite{16,17}, in order to overcome difficulties
from the CAST \cite{18} experiment, which represents an
alternative to the axion-like scenario. It is the so-called
Chern-Simons-like coupling scenario, which includes a light massive
vector field interacting with the familiar $U(1)_{QED}$ photon via a
Chern-Simons-like coupling. As a result, it was argued that this new
model reproduces the effects of rotation of the polarization plane.
In other terms, on the one hand both the Chern-Simons-like coupling
and axionic electrodynamics models are quite different and, on the
other hand, have optical-like features which they share. Therefore,
is rather justifiable to have some additional understanding of the
physical consequences presented by this new scenario (Chern-Simons-like
coupling scenario), from a somewhat different perspective.

Our goal in this letter is precisely to examine the impact of a light
massive vector field in the Chern-Simons-like coupling scenario through
a proper study of the concepts of screening and confinement. In this
perspective, we recall that the effective quantum
potential between two static charges is a key tool to study the equivalence
among models, which, otherwise are only suggested by other, very formal
approaches. Accordingly, our analysis reveals both expected and unexpected
features of the model under consideration. To accomplish our analysis, we
use the gauge-invariant but path-dependent variables formalism along the
lines of Ref. ${19-22}$, which is a physically-based alternative to the
usual Wilson loop approach and a preliminary version of this work has
appeared before \cite{23}.

\section{Interaction energy}

As mentioned above, the gauge theory we are considering describes the
interaction between the familiar massive $U(1)_{QED}$ photon with a
light massive vector field via a Chern-Simons-like coupling. In
this case the corresponding theory is governed by the Lagrangian
density \cite{16,17} :
\begin{eqnarray}
\mathcal{L} =  - \frac{1}{4}F_{\mu \nu }^2 (A) - \frac{1}{4}F_{\mu
\nu }^2 (B) + \frac{{m_\gamma ^2 }}{2}A_\mu ^2  + \frac{{m_B^2
}}{2}B_\mu ^2 - \frac{{\kappa }}{2}\varepsilon ^{\mu
\nu \lambda \rho } A_\mu  B_\nu F_{\lambda \rho } (A),
\label{Csmag05}
\end{eqnarray}
where $m_\gamma$ is the mass of the photon, and $m_{B}$ represents
the mass for the gauge boson $B$. Notice that this alternative
theory exhibits an effective mass for the component of the photon
along the direction of the external magnetic field, exactly as it
happens with axionic electrodynamics. Nevertheless, the Lagrangian
above is not so a-priori as Eq. (\ref{Csmag05}) seems to indicate.
Actually, it comes out as the unitary-gauge-fixed version of a more
fundamental and full gauge invariant action thoroughly discussed in
the works \cite{16} and \cite{17}.

Next, if we consider the model in the limit of a
very heavy $B$-field ($m_{B} \gg m_\gamma$) and we are bound to
energies much below $m_{B}$, we are allowed to integrate over
$B_\mu$ and to speak about an effective model for the $A_\mu$-field.
This can be readily accomplished by means of the path integral
formulation of the generating functional associated to
Eq.(\ref{Csmag05}), where the $B_\mu$-field appears at most
quadratically. By then shifting it according to the expression
\begin{equation}
B_\mu   \equiv \tilde B_\mu   + \frac{\kappa }{2}\frac{1} {{\left(
{\Delta  + m_B^2 } \right)}}\left[ {\eta _{\mu \nu } +
\frac{{\partial _\mu  \partial _\nu  }}{{m_B^2 }}}
\right]\varepsilon ^{\mu \lambda \rho \nu } A_\mu  F_{\lambda \rho }
\left( A \right), \label{Csmag10}
\end{equation}
and carrying out the Gaussian integration over the $\tilde
B_\mu$-field, we are lead to the effective Lagrangian for $A_\mu$,
as it follows below:
\begin{eqnarray}
\mathcal{L}_{eff}  &=&  - \frac{1}{4}F_{\mu \nu }^2  +
\frac{{m_\gamma ^2 }} {2}A_\mu  A^\mu   + \frac{{\kappa ^2
}}{4}A_\alpha  F_{\beta \gamma } \frac{1}
{{\left( {\Delta  + m_B^2 } \right)}}A^\alpha  F^{\beta \gamma } \nonumber\\
&+& \frac{{\kappa ^2 }}{2}A_\alpha  F_{\beta \gamma }
\frac{1}{{\left( {\Delta  + m_B^2 }
\right)}}A^\gamma  F^{\alpha \beta }
+ \frac{{\kappa ^2 }}{8}\tilde F^{\alpha \beta } F_{\alpha \beta }
\frac{1} {{m_B^2 \left( {\Delta  + m_B^2 } \right)}}\tilde F^{\gamma
\delta } F_{\gamma \delta }, \label{Csmag15}
\end{eqnarray}\\
where ${\widetilde F}_{\mu \nu }  \equiv {\raise0.7ex\hbox{$1$}
\!\mathord{\left/{\vphantom {1 2}}\right.\kern-\nulldelimiterspace}
\!\lower0.7ex\hbox{$2$}}\varepsilon _{\mu \nu \lambda \rho }
F^{\lambda \rho }$. One should notice that no gauge ambiguity is
present in the process of integrating out $B_\mu$, for
Eq.(\ref{Csmag05}) is already a gauge-fixed Lagrangian, as we have
pointed out in connection to Ref. ${16}$. Also, we observe
that the system described by the Lagrangian (\ref{Csmag15})
is a system with non-local time derivatives. However, we stress
that this paper is aimed at studying the static potential of
the theory (\ref{Csmag15}), so that $\Delta$ can be replaced by
$ - \nabla ^2$. For notational convenience we have
maintained $\Delta$, but it should be borne in mind that this paper
essentially deals with the static case. Thus, the canonical
quantization of this theory from the Hamiltonian point of view
follows straightforwardly, as we will show it below.

Now, if we wish to study quantum properties of the electromagnetic
field in the presence of external electric and magnetic fields, we
should split the $A_\mu$-field as the sum of a classical background,
$\langle A_\mu \rangle$, and a small quantum fluctuation, $a_\mu$,
namely: $A_\mu = \langle A_\mu \rangle + a_\mu$. Therefore the
previous Lagrangian density, up to quadratic terms in the fluctuations,
is also expressed as
\begin{eqnarray}
{\cal L}_{eff}  &=&  - \frac{1}{4}f_{\mu \nu } \Omega f^{\mu \nu} +
\frac{1}{2}a_\mu  M^2 a^\mu -\frac{{\kappa ^2 }}{2}f_{\gamma \beta
}\left\langle {A^\gamma  } \right\rangle \frac{{1}}{{\left( {\Delta
+ m_B^2 } \right)}} \left\langle {A_\alpha  } \right\rangle
f^{\alpha \beta } \nonumber\\
&+& \frac{{\kappa ^2 }}{8}f_{\mu \nu } v^{\mu \nu }
\frac{1} {{m_B^2 \left( {\Delta  + m_B^2 } \right)}}v^{\lambda \rho
} f_{\lambda \rho} - \frac{{\kappa ^2 }}{2}\left( {\varepsilon ^{jk0i} v_{0i}
\left\langle {A^l } \right\rangle a_l \frac{1}{{\left( {\Delta  +
m_B^2 } \right)}}f_{jk} } \right) \nonumber\\
&-& \kappa ^2 \left( {\varepsilon ^{jk0i} v_{0i} \left\langle {A_j }
\right\rangle a^m \frac{1}{{\left( {\Delta  + m_B^2 }
\right)}}f_{km} }
\right) + \kappa ^2 \left( {\varepsilon ^{jk0i} v_{0i} \left\langle {A^m }
\right\rangle a_k \frac{1}{{\left( {\Delta  + m_B^2 }
\right)}}f_{jm} }
\right), \nonumber\\
\label{Csmag25}
\end{eqnarray}
where $f_{\mu \nu } = \partial _\mu  a_\nu   - \partial _\nu a_\mu$,
and $\Delta \equiv \partial_\mu\partial^\mu$. $\Omega  \equiv 1 -
\kappa ^2 \frac{{\left\langle {A^i } \right\rangle \left\langle {A_i
} \right\rangle }}{{\left( {\Delta  + m_B^2 } \right)}}$, and $M^2
\equiv m_\gamma ^2  + \frac{{\kappa ^2 }}{2}\frac{{v_{0i} v^{0i}
}}{{\left( {\Delta  + m_B^2 } \right)}}$. In the above Lagrangian we
have considered the $v^{0i}\neq0$ and $v^{ij}=0$ case (referred to
as the magnetic one in what follows), and simplified our notation by
setting $\varepsilon ^{\mu \nu \alpha \beta } \left\langle{F_{\mu
\nu } } \right\rangle  \equiv v^{\alpha \beta }$. As a result,
the Lagrangian (\ref{Csmag25}) becomes a Maxwell-Proca-like theory
with a manifestly Lorentz violating term.

This effective theory provide us with a suitable starting point to
study the interaction energy. However, before proceeding with the
determination of the energy, it is necessary to restore the gauge
invariance in (\ref{Csmag25}). Standard techniques for constrained
systems then lead to the following effective Lagrangian:
 \begin{eqnarray}
\mathcal{L}_{eff}  &=&  - \frac{1}{4}f_{\mu \nu } \left[
{\frac{{\left( {\Delta ^2 + a^2 \Delta  + b^2 } \right)}}{{\Delta
\left( {\Delta  + m_B^2 } \right)}}} \right]f^{\mu \nu }
-\left\langle {A^i } \right\rangle f_{i0} \frac{{1}}
{{\left( {\Delta + m_B^2 } \right)}} \left\langle {A_k }
\right\rangle f^{k0} \nonumber\\
&-& \frac{{\kappa ^2 }}{2}f_{ki} \left\langle {A^k } \right\rangle
\frac{{1}} {{\left( {\Delta  + m_B^2 } \right)}} \left\langle {A_l }
\right\rangle f^{li}
+ \frac{{\kappa ^2 }}{8}v^{0i}
f_{0i} \frac{1}{{m_B^2 \left( {\Delta  + m_B^2 } \right)}}v^{0k}
f_{0k}, \label{Csmag40}
\end{eqnarray}
where $a^2  \equiv m_B^2  + m_\gamma ^2 \left( {1 - \kappa ^2
\frac{{\left\langle {A_k  } \right\rangle \left\langle {A^k  }
\right\rangle }} {{m_\gamma ^2 }}} \right)$, and $ b^2  = m_\gamma
^2 \left( {m_B^2  - \frac{{\kappa ^2 {\bf v}^2 }}{{2m_\gamma ^2 }}}
 \right)$. To get the above theory we have ignored
the last three terms in (\ref{Csmag25}) because it add nothing to the
static potential calculation, as we will show it below. Consequently,
the new effective action (\ref{Csmag40}) provide us with a suitable
starting point to study the interaction energy without loss of
physical content.

Having established the new effective Lagrangian, we can now compute
the interaction energy. To this end, we first consider the Hamiltonian
framework of this new effective theory. The canonical momenta
read $\Pi ^\mu   =  - \left( {\frac{{\Delta ^2  + a^2 \Delta  + b^2
}} {{\Delta \left( {\Delta  + m_B^2 } \right)}}} \right)f^{0\mu } +
\kappa ^2 \frac{{\left\langle {A^\mu } \right\rangle \left\langle
{A_k } \right\rangle }}{{\left( {\Delta + m_B^2 } \right)}}f^{k0}
+ \frac{{\kappa ^2 }}{4}\frac{{v^{0\mu } }}{{m_B^2 }}\frac{1}
{{\left( {\Delta  + m_B^2 } \right)}}v^{0k} f_{0k}$, and one
immediately identifies the usual primary constraint $\Pi ^i  =  -
\left( {\frac{{\Delta ^2  + a^2 \Delta  + b^2 }} {{\Delta \left(
{\Delta  + m_B^2 } \right)}}} \right)f^{0i} + \kappa ^2
\frac{{\left\langle {A^i } \right\rangle \left\langle {A_k }
\right\rangle }}{{\left( {\Delta  + m_B^2 } \right)}}f^{k0} +
\frac{{\kappa ^2 }}{4}\frac{{v^{0i} }}{{m_B^2 }}\frac{1} {{\left(
{\Delta  + m_B^2 } \right)}}v^{0k} f_{0k}$. Therefore the canonical
Hamiltonian takes the form
\begin{eqnarray}
H_C  &=& \int {d^3 x} \left\{ { - a_0 \partial _i \Pi ^i  +
\frac{1}{2}B^i \frac{{\left( {\Delta ^2  + a^2 \Delta + b^2 }
\right)}}{{\Delta \left( {\Delta  + m_B^2 } \right)}}B^i } \right\}
- \frac{1}{2}\int {d^3 } x\Pi _i \frac{{\left( {\Delta  + m_B^2 }
\right)}}
{{\left( {\Delta  + a^2  + \frac{{b^2 }}{\Delta }} \right)}} \Pi^i \nonumber\\
&+& \frac{{\kappa ^2 }}{2}\int {d^3 x} \Pi _i \left\langle {A^i }
\right\rangle \frac{{\left( {\Delta  + m_B^2 } \right)}}{{\left(
{\Delta  + a^2  + \frac{{b^2 }}{\Delta }} \right)^2}}\left\langle
{A^k } \right\rangle \Pi _k
+ \frac{{\kappa ^2 }}{2}\int {d^3 x} f_{ki} \left\langle {A^k }
\right\rangle \frac{1}{{\left( {\Delta  + m_B^2 }
\right)}}\left\langle {A_l } \right\rangle f^{li} \nonumber\\
&+& \frac{{\kappa ^2 {\bf v}^2 }}{{8m_B^2 }}\int {d^3 x} \Pi _i
\frac{{\left( {\Delta + m_B^2 } \right)}}{{\left( {\Delta  + a^2  +
\frac{{b^2 }}{\Delta }} \right)^2 }}\Pi ^i  ,\label{Csmag45}
\end{eqnarray}
where $a^2  = m_B^2  + m_\gamma ^2  + \kappa ^2 \left\langle {\bf A}
\right\rangle ^2$ and $b^2  = m_\gamma ^2 m_B^2  + \frac{{\kappa ^2
}}{2}{\bf v}^2$. Since our energies are all much below $m_B$, it is
consistent with our considerations to neglect $\kappa ^2
\left\langle {\bf A} \right\rangle ^2$ with respect to $m_B^2$. This
implies that $a^2$ and $b^2$ should be taken as: $a^2  = m_B^2$ and
$b^2  = m_\gamma ^2 m_B^2  + \frac{{\kappa ^2 }}{2}{\bf v}^2$.

Temporal conservation of the primary constarint $\Pi _0$ leads to
the secondary constraint $\Gamma_1 \left( x \right) \equiv \partial _i \Pi
^i=0$. It is also possible to verify that no further constraints are
generated by this theory. The extended Hamiltonian
that generates translations in time then reads $H = H_C + \int
{d^2}x\left( {c_0 \left( x \right)\Pi _0 \left( x \right)
 + c_1 \left(x\right)\Gamma _1 \left( x \right)} \right)$, where $c_0 \left(
x\right)$ and $c_1 \left( x \right)$ are arbitrary Lagrange
multipliers. Moreover, it follows from this Hamiltonian that
$\dot{a}_0 \left( x \right)= \left[ {a_0 \left( x \right),H} \right]
= c_0 \left( x \right)$, which is completely arbitrary. Since $
\Pi^0 = 0$ always, neither $a^0$ and $\Pi^0$ are of interest in
describing the system and may be discarded from the theory. If a new
arbitrary coefficient $c(x) = c_1 (x) - A_0 (x)$ is introduced the
Hamiltonian may be rewritten as
\begin{eqnarray}
H  &=& \int {d^3 x} \left\{ {c(x)\partial _i \Pi ^i  + \frac{1}{2}B^i
\frac{{\left( {\Delta ^2  + a^2 \Delta  + b^2 } \right)}}{{\Delta
\left( {\Delta  + m_B^2 } \right)}}B^i } \right\} - \frac{1}{2}\int {d^3 } x\Pi _i
\frac{{\left( {\Delta  + m_B^2 } \right)}}{{\left( {\Delta  + a^2  + \frac{{b^2 }}
{\Delta }} \right)}}\Pi^i \nonumber\\
&+& \frac{{\kappa ^2 }}{2}\int {d^3 x} \Pi _i \left\langle {A^i } \right\rangle
\frac{{\left( {\Delta  + m_B^2 } \right)}}{{\left( {\Delta  + a^2  + \frac{{b^2 }}
{\Delta }} \right)^2}}\left\langle {A^k } \right\rangle \Pi _k
+ \frac{{\kappa ^2 }} {2}\int {d^3 x} f_{ki} \left\langle {A^k } \right\rangle \frac{1}
{{\left( {\Delta  + m_B^2 } \right)}}\left\langle {A_l } \right\rangle f^{li}
\nonumber\\
&+& \frac{{\kappa ^2 {\bf v}^2 }}{{8m_B^2 }}\int {d^3 x} \Pi _i
\frac{{\left( {\Delta  + m_B^2 } \right)}}{{\left( {\Delta  + a^2  +
\frac{{b^2 }}{\Delta }} \right)^2 }}\Pi ^i. \label{Csmag50}
\end{eqnarray}

Since there is one first class constraint $\Gamma_{1}(x)$ (Gauss' law),
we choose one gauge fixing condition that will make the full set of
constraints becomes second class. We choose the gauge fixing condition
to correspond to
\begin{equation}
\Gamma _2 \left( x \right) \equiv \int\limits_{C_{\xi x} } {dz^\nu }
a_\nu \left( z \right) \equiv \int\limits_0^1 {d\lambda x^i } a_i
\left( {\lambda x} \right) = 0,     \label{Csmag55}
\end{equation}
where  $\lambda$ $(0\leq \lambda\leq1)$ is the parameter describing
the spacelike straight path $ x^i = \xi ^i  + \lambda \left( {x -
\xi } \right)^i $, and $ \xi $ is a fixed point (reference point).
There is no essential loss of generality if we restrict our
considerations to $ \xi ^i=0 $. The choice (\ref{Csmag55}) leads to
the Poincar\'e gauge \cite{25,26}. As a consequence, we can now
write down the only nonvanishing Dirac bracket for the canonical
variables
\begin{eqnarray}
\left\{ {a_i \left( x \right),\Pi ^j \left( y \right)} \right\}^ *
&=&\delta{ _i^j} \delta ^{\left( 3 \right)} \left( {x - y} \right)
- \partial _i^x \int\limits_0^1 {d\lambda x^j } \delta ^{\left( 3
\right)} \left( {\lambda x - y} \right). \label{Csmag60}
\end{eqnarray}

We pass now to the calculation of the interaction energy, where
a fermion is localized at the origin $\bf 0$ and an antifermion at
$\bf y$. For this purpose, we will calculate the
expectation value of the energy operator $H$ in the physical state
$|\Phi\rangle$. In this context, we recall that the physical state
$|\Phi\rangle$ can be written as
\begin{eqnarray}
\left| \Phi  \right\rangle \equiv   \left| {\overline \Psi
\left(
\bf y \right)\Psi \left( {\bf 0} \right)} \right\rangle
= \overline \psi \left( \bf y \right)\exp \left(
{iq\int\limits_{{\bf 0}}^{\bf y} {dz^i } a_i \left( z \right)}
\right)\psi \left({\bf 0} \right)\left| 0 \right\rangle,
\label{Csmag65}
\end{eqnarray}
where $\left| 0 \right\rangle$ is the physical vacuum state. The
line integral is along a spacelike path starting at $\bf 0$ and
ending at $\bf y$, on a fixed time slice.

Returning now to our problem on hand, we compute the expectation value
of $H$ in the physical state $|\Phi\rangle$. Taking into account the above
Hamiltonian structure, we observe that
\begin{eqnarray}
\Pi _i \left( x \right)\left| {\overline \Psi  \left( \bf y
\right)\Psi \left( {{\bf y}^ \prime  } \right)} \right\rangle  =
\overline \Psi  \left( \bf y \right)\Psi \left( {{\bf y}^ \prime }
\right)\Pi _i \left( x \right)\left| 0 \right\rangle
+ q\int_ {\bf y}^{{\bf y}^ \prime  } {dz_i \delta ^{\left( 3
\right)} \left( {\bf z - \bf x} \right)} \left| \Phi
\right\rangle.
\label{Csmag65b}
\end{eqnarray}
Having made this observation and since the fermions are taken to be
infinitely massive (static sources), we can substitute $\Delta$ by
$-\nabla^{2}$ in Eq. (\ref{Csmag50}). Therefore, the expectation
value $\left\langle H \right\rangle _\Phi$ is expressed as
\begin{equation}
\left\langle H \right\rangle _\Phi   = \left\langle H \right\rangle
_0 + \left\langle H \right\rangle _\Phi ^{\left( 1 \right)} +
\left\langle H \right\rangle _\Phi ^{\left( 2 \right)},
\label{Csmag70}
\end{equation}
where $\left\langle H \right\rangle _0  = \left\langle 0
\right|H\left| 0 \right\rangle$. The $\left\langle H \right\rangle
_\Phi ^{\left( 1 \right)}$ and $\left\langle H \right\rangle _\Phi
^{\left( 2 \right)}$ terms are given by
\begin{eqnarray}
\left\langle H \right\rangle _\Phi ^{\left( 1 \right)}  =  -
\frac{{b^2 B}}{2} \int {d^3 x} \left\langle \Phi  \right|\Pi _i
\frac{{\nabla ^2 }} {{\left( {\nabla ^2  - M_1^2 } \right)}}\Pi ^i
\left| \Phi  \right\rangle
+\frac{{b^2 B}}{2}\int {d^3 x}
\left\langle \Phi  \right|\Pi _i \frac{{M_2^2 }} {{M_1^2
}}\frac{{\nabla ^2 }}{{\left( {\nabla ^2  -
M_2^2 } \right)}} \Pi ^i \left| \Phi  \right\rangle,
\label{Csmag75a}
\end{eqnarray}
and
\begin{eqnarray}
\left\langle H \right\rangle _\Phi ^{\left( 2 \right)}  = m_B^2
b^2 B\int {d^3 x} \left\langle \Phi  \right|\Pi _i \frac{1}{{\left(
{\nabla ^2  - M_1^2 } \right)}} \Pi ^i \left| \Phi  \right\rangle
- m_B^2 b^2 B\int {d^3 x} \left\langle \Phi  \right|\Pi _i
\frac{{M_2^2 }} {{M_1^2 }}\frac{1}{{\left( {\nabla ^2  - M_2^2 }
\right)}}\Pi ^i \left|
\Phi  \right\rangle,
\label{Csmag75b}
\end{eqnarray}
where $B = \frac{1}{{M_2^2 \left( {M_2^2  - M_1^2 } \right)}}$,
$M_1^2  \equiv \frac{{a^2 }}{2}\left[ {1 + \sqrt {1 - \frac{{4b^2 }}
{{a^4 }}} } \right]$, and $M_2^2  \equiv \frac{{a^2 }} {2}\left[ {1
- \sqrt {1 - \frac{{4b^2 }}{{a^4 }}} } \right]$.

We have neglected the terms in (\ref{Csmag50}) where $\left( {\Delta
+ a^2  + \frac{{b^2 }}{\Delta }} \right)^2$ appears in the
denominator, the reason being that we wish to compute an
interparticle potential, which expresses the effects of photons
exchange in the low-energy (or low-frequency) limit. Therefore,
these terms we are mentioning are suppressed in view of the presence
of higher power of the frequency in the denominator. Another
important point to be highlighted in our discussion comes from the
expressions for $M_1^2$ and $M_2^2$. Our treatment is only valid
under the assumption that $a^4 > 4b^2$. However, this condition is
equivalent to taking $\kappa ^2 {\bf v}^2  < \frac{{m_B^4 }}{2}$,
which is perfectly compatible with our approximation. So, we
restrict ourselves to an external magnetic field such that $|{\bf
v}| < \frac{{m_B^2 }}{{2\kappa ^2 }}$.

Following our earlier procedure $^{19-24}$, we see
that the potential for two opposite charges located at ${\bf 0}$ and
${\bf y}$ takes the form
\begin{eqnarray}
V &=&  - \frac{{q^2 }}{{4\pi }} \Biggl[\frac{1}{2}\left( {1 - \frac{3}{{\sqrt {1 - 4\left( {{\raise0.5ex\hbox{$\scriptstyle {m_\gamma ^2 }$}
\kern-0.1em/\kern-0.15em
\lower0.25ex\hbox{$\scriptstyle {m_B^2 }$}} + {\raise0.5ex\hbox{$\scriptstyle {2\kappa ^2 B^2 }$}
\kern-0.1em/\kern-0.15em
\lower0.25ex\hbox{$\scriptstyle {m_B^4 }$}}} \right)} }}} \right)\frac{1}{L}\exp \left[ {\left( { - \frac{{m_B }}{\sqrt 2}\sqrt {1 + \sqrt {1 - 4\left( {{\raise0.5ex\hbox{$\scriptstyle {m_\gamma ^2 }$}
\kern-0.1em/\kern-0.15em
\lower0.25ex\hbox{$\scriptstyle {m_B^2 }$}} + {\raise0.5ex\hbox{$\scriptstyle {2\kappa ^2 B^2 }$}
\kern-0.1em/\kern-0.15em
\lower0.25ex\hbox{$\scriptstyle {m_B^4 }$}}} \right)} } } \right)L} \right] \nonumber\\ 
&+& \frac{1}{2}\left( {1 + \frac{3}{{\sqrt {1 - 4\left( {{\raise0.5ex\hbox{$\scriptstyle {m_\gamma ^2 }$}
\kern-0.1em/\kern-0.15em
\lower0.25ex\hbox{$\scriptstyle {m_B^2 }$}} + {\raise0.5ex\hbox{$\scriptstyle {2\kappa ^2 B^2 }$}
\kern-0.1em/\kern-0.15em
\lower0.25ex\hbox{$\scriptstyle {m_B^4 }$}}} \right)} }}} \right)\frac{1}{L}\exp \left[ {\left( { - \frac{{m_B }}{\sqrt 2}\sqrt {1 - \sqrt {1 - 4\left( {{\raise0.5ex\hbox{$\scriptstyle {m_\gamma ^2 }$}
\kern-0.1em/\kern-0.15em
\lower0.25ex\hbox{$\scriptstyle {m_B^2 }$}} + {\raise0.5ex\hbox{$\scriptstyle {2\kappa ^2 B^2 }$}
\kern-0.1em/\kern-0.15em
\lower0.25ex\hbox{$\scriptstyle {m_B^4 }$}}} \right)} } } \right)L} \right] \Biggr]\ .
\label{Csmag85}
\end{eqnarray}  
Here ${\cal B}$ represents the external magnetic field and 
$|{\bf y}| \equiv L$. This result immediately shows that the theory
under consideration describes an exactly screening phase. It is
worthwhile stressing that the choice of the gauge is in this 
development really arbitrary. This then implies that we would obtain
exactly the same result in any gauge. It is appropriate to observe here
that by considering the limit $m_\gamma$  and  $\kappa$  $\to 0$, we obtain a theory of two independent uncoupled $U(1)$ gauge bosons, one of which is massless. In such a case, one can easily verify that the static potential is a Yukawa-like correction to the usual static Coulomb potential.

\section{Final Remarks}

In summary, we have considered the confinement versus screening issue
for a gauge theory which includes a light massive vector field interacting
with the familiar $U(1)_{QED}$ photon via a Chern-Simons- like coupling.
Interestingly enough, expression (\ref{Csmag85}) displays a marked departure
from the result of axionic electrodynamics. As already expressed, axionic
electrodynamics has a different structure which is reflected in a
confining piece, which is not present in the Chern-Simons-like
coupling scenario. Nevertheless, the above static potential profile is
similar to that encountered in axionic electrodynamics consisting of
a massless axion-like field. In fact, the linear confining
potential seems to be associated only with massive axion-like
particles. As opposed to non-Abelian axionic electrodynamics that,
even in the massless case, displays a confining potential. In addition,
we also mention that the above static potential profile is analogous to
that encountered in the coupling between the familiar massless
electromagnetism $U(1)_{QED}$ and a hidden-sector $U(1)_h$ inside a
superconducting box \cite{24}. In this connection it becomes of interest,
to recall that an experiment for searching for extra hidden-sector
$U(1)$ gauge bosons with small gauge kinetic mixing ($\chi$) with
the ordinary photon (in the laboratory) has been proposed in  \cite{13}.
Basically, this experiment consists in putting a sensitive magnetometer
inside a superconducting shielding, which in turn is placed inside a
strong magnetic field. In this case, it was argued that photon -
hidden photon oscillations would allow to penetrate the
superconductor and a magnetic field would register on the
magnetometer, in contrast with the usual electrodynamics where the
magnetic field cannot penetrate the superconductor. In
this setup the magnetometer measures the field strength which is
proportional to $\chi^2$. In this sense the present work could be of
interest for searching for bounds of the gauge kinetic mixing ($\chi$)
along the lines of expression (\ref{Csmag85}). Thus, we have
established a new connection between extensions of the Standard Model
(SM) such as axion-like particles and light extra $U(1)$ gauge bosons.
Accordingly, the benefit of considering the present approach is to provide
unifications among different models, as well as exploiting the
equivalence in explicit calculations, as we have illustrated in the
course of this work.

Finally, it should be noted that  by substituting $B_\mu$ by
$\partial_\mu \phi$ in (\ref{Csmag05}), the theory under consideration
assumes the form \cite{27}
\begin{equation}
{\cal L} =  - \frac{1}{4}F_{\mu \nu }^2  + \frac{{m_\gamma ^2 }}
{2}A_\mu ^2+ \frac{1}{2}\partial _\mu  \phi \partial ^\mu  \phi  - \frac{\kappa
}{{2m_B }}\varepsilon ^{\mu \nu \lambda \rho } F_{\mu \nu }
F_{\lambda \rho } \phi, \label{Csmag90}
\end{equation}
which is similar to axionic electrodynamics. In fact, it is worth
recalling here that axionic electrodynamics is described by
\cite{15}
\begin{equation}
{\cal L} =  - \frac{1}{4}F_{\mu \nu } F^{\mu \nu }  -
\frac{g}{8}\phi \varepsilon ^{\mu \nu \rho \sigma } F_{\mu \nu }
F_{\rho \sigma } + \frac{1}{2}\partial _\mu  \phi \partial ^\mu \phi
- \frac{{m_A^2 }}{2}\phi ^2, \label{Csmag90b}
\end{equation}
hence we see that both theories are quite different.

Next, after performing the integration over $\phi$ in (\ref{Csmag90}), the
effective Lagrangian density reads
\begin{eqnarray}
{\cal L} =  - \frac{1}{4}F_{\mu \nu }^2  + \frac{{m_\gamma ^2
}}{2}A_\mu ^2 - \frac{{\kappa ^2 }}{{8m_B^2 }}\varepsilon ^{\mu \nu
\lambda \rho } F_{\mu \nu } F_{\lambda \rho } \frac{1}{{\nabla ^2
}} \varepsilon ^{\alpha \beta \gamma \delta } F_{\alpha \beta }
F_{\gamma \delta }. \label{Csmag95}
\end{eqnarray}
This expression can now be rewritten as
\begin{eqnarray}
{\cal L} =  - \frac{1}{4}f_{\mu \nu }^2  + \frac{{m_\gamma ^2
}}{2}a_\mu ^2 - \frac{{\kappa ^2 }}{{2m_B^2 }}\varepsilon ^{\mu \nu
\alpha \beta } \left\langle {F_{\mu \nu } } \right\rangle
\varepsilon ^{\lambda \rho \gamma \delta } \left\langle {F_{\lambda
\rho } } \right\rangle f_{\alpha \beta } \frac{1}{{\nabla ^2
}}f_{\gamma \delta } , \label{Csmag100}
\end{eqnarray}
where $\left\langle {F_{\mu \nu } } \right\rangle$ represents the
constant classical background. Here $f_{\mu \nu } =\partial _\mu
a_\nu -\partial _\nu a_\mu$ describes fluctuations around the
background. The above Lagrangian arose after using $ \varepsilon
^{\mu \nu \alpha \beta } \left\langle {F_{\mu \nu } } \right\rangle
\left\langle {F_{\alpha \beta } } \right\rangle=0$ (which holds for
a pure electric or a pure magnetic background). By introducing the
notation $\varepsilon ^{\mu \nu \alpha \beta }
 \left\langle{F_{\mu \nu } } \right\rangle  \equiv v^{\alpha \beta }
 $ and $\varepsilon ^{\rho \sigma \gamma \delta } \left\langle {F_{\rho
\sigma } } \right\rangle  \equiv v^{\gamma \delta }$, expression
(\ref{Csmag100}) then becomes
\begin{equation}
{\cal L} =  - \frac{1}{4}f_{\mu \nu }^2  + \frac{{m_\gamma ^2
}}{2}a_\mu ^2 - \frac{{\kappa ^2 }}{{2m_B^2 }}v^{\alpha \beta }
f_{\alpha \beta } \frac{1}{{\nabla ^2 }}v^{\gamma \delta } f_{\gamma
\delta} , \label{Csmag105}
\end{equation}
where the tensor $v^{\alpha \beta }$ is not arbitrary, but satisfies
$\varepsilon ^{\mu \nu \alpha \beta } v_{\mu \nu } v_{\alpha \beta
}=0$. With this in view, we now proceed to calculate the interaction
energy in the $v^{0i}\ne0$ and $v^{ij}=0$ case (referred to as the magnetic one in what follows).

Following the same steps employed for obtaining (\ref{Csmag85}), the
static potential is expressed as
\begin{equation}
V =  - \frac{{q^2 }}{{4\pi }}\frac{1}{L}\exp \left[ { - \left( {\sqrt {m_\gamma ^2  + {\raise0.5ex\hbox{$\scriptstyle {4\kappa ^2 {\cal B}^2 }$}
\kern-0.1em/\kern-0.15em
\lower0.25ex\hbox{$\scriptstyle {m_B^2 }$}}} } \right)L} \right], 
\label{Csmag130}
\end{equation}
where $\cal B$ represents the external magnetic field. We observe
that in the limit $m_\gamma   \to 0$, axionic electrodynamics
experiences mass generation induced by an external magnetic field.
In this way the theory describes a screening phase, as we have just
seen above. Evidently, by considering the limit $\kappa \to
 0$, we obtain a Proca-like theory.

\section{Acknowledgments}

It is a pleasure to thank J. A. Hela\"{y}el-Neto and E. Spallucci for
collaboration and useful discussions. I would also like to thank the Field
Theory Group of the CBPF and the Physics Department  of the Universit\`a
di Trieste for hospitality. This work was supported in part by Fondecyt
(Chile) grant 1080260.

\end{document}